\documentclass{article}

\usepackage[margin=0.75in]{geometry}
\usepackage{graphicx,graphics}
\usepackage{amsmath,amsfonts}
\usepackage{calligra}
\DeclareMathAlphabet{\mathcalligra}{T1}{calligra}{m}{n}

\newcommand{\bfb} {\mbox{\boldmath $\omega$}}

\newcommand{\x}{{\mbox{\boldmath $x$}}}
\newcommand{\X}{{\mbox{\boldmath $X$}}}
\newcommand{\y}{{\mbox{\boldmath $y$}}}

\newcommand{\q}{{\mbox{\boldmath $q$}}}
\newcommand{\I}{{\mbox{\boldmath $I$}}}

\newtheorem{theorem}{Theorem}[section]
\newtheorem{lemma}{Lemma}[section]
\newtheorem{assumption}{Assumption}[section]

\title{\textbf{Regularization and Variable Selection with Copula Prior}}

\author{Rahul~Sharma
      ~~  and ~~ Sourish~Das
\thanks{Rahul~Sharma is with the Department of Humanities and Social
  Sciences (Economics), Indian Institute of Technology Kanpur, India
  e-mail: rahul2057210 at gmail dot com}
\thanks{Sourish Das is with the Chennai Mathematical Institute, India
  e-mail: sourish at cmi dot ac do in}
}

\begin{document}

\maketitle

\begin{abstract}
In this work, we show that under specific choices of the copula, the lasso,
elastic net, and $g$-prior are particular cases of `copula prior,' for
regularization and variable selection method. We present `lasso with Gauss
copula prior' and `lasso with t-copula prior.' The simulation study and
real-world data for regression, classification, and large time-series data
show that the `copula prior' often outperforms the lasso and elastic net
while having a comparable sparsity of representation. Also, the copula
prior encourages a grouping effect. The strongly correlated predictors tend
to be in or out of the model collectively under the copula prior. The
`copula prior' is a generic method, which can be used to define the new
prior distribution. The application of copulas in modeling prior
distribution for Bayesian methodology has not been explored much. We
present the resampling-based optimization procedure to handle big data with
copula prior.
\end{abstract}


\section{Introduction}\label{submission}

A machine learning algorithm can perform supervised learning task,
using a set of features \cite{Guyon.2003,
  Chandrashekar.2014}. Variable Selection helps reducing computation
requirement, reducing the effect of `curse of dimensionality,' improve
the prediction performance and reveal the relationship between
predictors and the target variable. In microarray data which usually
consists of the `expression state' of a vast number of genes, it is
extremely desirable to pick multiple correlated genes to reveal
insights into the underlying biological pathway. Selecting correlated
variables often presents a challenge to the classical variable
selection methods. `Lasso' proposed by \cite{Tibshirani.1996} is a
popular choice for variable selection. It uses a $l_{1}$ penalty on
the model parameters. However, lasso selects only a small subset of
variables, from a group of highly correlated variables; affects the
prediction accuracy as well as the interpretability of the estimated
model.

To address this problem, \cite{Zou.Hastie.2005} proposed `elastic net'
(EN), which encourage a grouping effect; where strongly correlated variables tend to be in or out of the model together.  However, the EN  prior distribution is simple, and it does not incorporate correlation
information among variables in the model. To fix this issue, several
other regularizers have been developed. \cite{Peter.2013} describes a two-stage process in which one first cluster the features to identify the correlated variables and then apply lasso type penalties to learn the model. But now attempts are made to avoid the two-stage process and use a regularizer which could simultaneously
learn the coefficients and can identify the groups of strongly
correlated variables. Ordered weight $l_{1}$ (OWL) devised by
\cite{Zeng.2016} can discover the groups of strongly correlated
variables. However, OWL usually forces the coefficients within the
same group to have the similar value which makes it
undesirable. Another useful feature selection algorithm in this
context is the eigennet \cite{Yan.2011}. It selects the correlated variables by using the eigenstructure of the data to guide feature selection. From a Bayesian perspective, natural way to deal with the problem is to use multivariate Laplace distribution as a regularizer, which can account for the correlation between the coefficients, like a $g$-prior developed by \cite{Zellner.1986}.  However, the multivariate Laplace distribution is complicated, as its pdf involves the modified Bessel function of the second kind \cite{Kotz.2001}. So computationally it becomes difficult to handle.

We present the multivariate version of the lasso, called the `lasso copula'
(LC) prior which can incorporate the correlation information between the
features.  Due to its built-in correlation structure, it can discover the
groups of strongly correlated variables. The advantage of our proposed LC
prior is that it encourages grouping effect with an appropriate sparsity of
representation. The LC prior just like the lasso or EN can perform both
feature selection as well as regularization.

For estimating the coefficients, we propose a nonlinear optimization
procedure and resampling-based optimization procedure to handle big data. Through experiments on simulated data and real-life data sets, we show that the LC prior can outperform the state of the
art methods like the regular lasso and EN.

\subsection{Contribution}
\begin{itemize}
\item We present the `lasso Gauss copula' (LGC) prior $\&$ `lasso
  $t$-copula' (LTC) prior
  which can use the correlation information embedded in the data
  to select correlated variables.

\item We show that LGC reduces to regular lasso prior when the
  correlation between the features is 0. Hence understanding the
  theoretical properties of LGC prior is of significant interest.

\item We propose a framework for tuning the hyperparameters of LC
    prior. For estimating the coefficients, a non-linear optimization procedure is employed, and resampling procedure is
    presented to handle large dataset with large feature space.
\end{itemize}

\section{Proposed Method}
We first describe the problem statement in detail, then
introduce the LC prior.

\subsection{Problem Statement}
Consider a linear regression problem consisting of $n$ independent and
identically distributed samples $\{x_{i},y_{i}\}_{i=1}^{n}$.   Here
$x_{i}$ denotes a $p$ dimensional input vector for which output value
is denoted by $y_{i}$. Although we consider the regression problem
here;  later we extended our approach to the classification and
time-series problem. For every sample $i$, we  implement the following model
\begin{equation}
y_{i}=x_{i}^T\bfb+\epsilon_{i},~~~i=1,2,...,n,
\end{equation}
where $\epsilon_{i}\sim \mathcal{N}(0, \sigma^{2})$. Our goal is to select the correct set of features $\&$
learn the true value of coefficient vector  $\bfb \in \mathbb{R}^p$ which
relates $y_{i}$ and $x_{i}$. For estimating $\bfb$, we minimize the
squared error loss function with LC prior as the regularizer. We define
$\y=(y_{1},\ldots,y_{n})$ to be $n \times 1$ column vector of responses and
$\x=(x_{1},\ldots,x_{n})^{T}$ as an $n \times p$ matrix of
features. \textit{Without loss of generality, we
assume that each response has been centered and each predictor has been
standardized.}

\subsection{Copula Prior}
Joint modeling of variables could be complicated if the marginals are not
Gaussian, i.e., it belongs to different parametric families. In such cases,
we can use copula techniques to define the multivariate distribution
functions. A copula is a function that connects univariate marginals to their
full multivariate distribution. The application of Copulas in modeling priors has not been explored much. In this
paper, we present how copula can be used to develop the joint priors over the
parameters.

Mathematically copula can be defined as a $p$ dimensional function $C$,
$$
C:[0,1]^p\rightarrow [0,1].
$$
The  Sklar's theorem \cite{Sklar.1959} states that every multivariate
distribution function can be expressed as
\begin{equation}
\label{eqn_Sklar_copula}
F(\bfb)=C(F_{1}(\omega_{1}),\ldots,F_{p}(\omega_{p}),\theta),
\end{equation}
where $\theta$ is the dependence parameter and $F_i(\omega),~i=1,2,\ldots,p$
are marginal prior distributions. If
$F_{1}(\omega_{1}),\ldots,F_{p}(\omega_{p})$ are continuous then  $\exists$
an unique $C$ satisfying (\ref{eqn_Sklar_copula}). If we consider the product
copula, i.e.,
\begin{equation}
\label{eqn_product_copula}
C(u_1,\ldots,u_p)=u_1\ldots u_p,
\end{equation}
and choose Gaussian distribution over $\omega_j$, i.e.,
$u_j=F_j(\omega_j)=\Phi(\omega_j,0,\tau)$ as marginal prior distribution then
it is ridge prior and corresponding penalty is $L_2$ penalty. If we choose
Laplace distribution as marginals, i.e.,
$F_j(\omega_j)=Laplace(\omega_j,0,\tau)$ and consider the product copula as
(\ref{eqn_product_copula}) then it is lasso prior  and corresponding penalty
is $L_1$ penalty. Similarly if we choose EN distribution over $\omega_j$ as
marginal prior distribution and consider the product copula
(\ref{eqn_product_copula}), then it is EN prior and the corresponding penalty
is the convex combination between $L_1$ and $L_2$-norm. Following the similar
argument, if we choose the marginal distribution to be Gaussian distribution
and consider Gaussian copula with covariance matrix to be $\Sigma =
g(X^TX)^{-1}$, then it is $g$-prior \cite{Zellner.1986}. As it turns out,
the existing priors like ridge, lasso, EN, $g$-priors becomes special cases
of the proposed copula prior, for the particular choices of copula. We
present the complete list of existing cases and new copula priors in the
table \ref{table_list_of_copula_priors}.

\begin{table*}[ht]
\centering
\begin{tabular}{cccc}\hline
Marginal Distribution & Copula Type & Covariance & Prior \\ \hline\hline
Normal  & product copula & $\I$ &ridge \\
Laplace & product copula & $\I$ &lasso \\
Elastic Net & product copula & $\I$ &elastic net\\
Normal  & Multivariate Gaussian  & $g(X^TX)^{-1}$ & $g$-prior\\
Laplace & Multivariate Gaussian  & $\Sigma$ & lasso-Gauss-Copula\\
Laplace & Multivariate $t$ with $\nu$ df & $\Sigma$ & lasso-$t$-Copula\\
Laplace & Multivariate Cauchy & $\Sigma$ & lasso-Cauchy-Copula\\
\hline
\end{tabular}
\caption{List of Copula Prior for Regularizations, $\I$ is identity
  matrix and $\Sigma$ is the unstructured/structured covariance matrix
needs to be estimated.}
\label{table_list_of_copula_priors}
\end{table*}

As \cite{Sklar.1959} showed that a multivariate distribution can be written
as a Copula,
\begin{equation}
\label{eqn_CDF_copula}
F[F_{1}^{-1}(u_{1}),\ldots,F_{p}^{-1}(u_{p})]=C(u_{1},\ldots,u_{p},\Sigma),
\end{equation}
where $F_{j}^{-1}(u_{j})=\omega_j,~j=1\ldots p$. Now if we consider the Gauss
copula, as $F=\Phi$, differentiating the equation (\ref{eqn_CDF_copula}) with
respect to $u_{1},\ldots, u_{p}$, we get the derivative of copula as
\begin{equation}\label{eqn_copula_derivative}
c(u_{1},\ldots,u_{p};\Sigma)=\frac{f[F_{1}^{-1}(u_{1}),\ldots,F_{p}^{-1}(u_{p})]}{\prod_{i=1}^{p}f_{i}[F_{i}^{-1}(u_{i})]}.
\end{equation}
The $f$ in (\ref{eqn_copula_derivative}) is the joint PDF of the $F$, and
$f_1,f_2,\ldots,f_p$ are univariate marginal density functions. The
expression (\ref{eqn_copula_derivative}) holds for any choice of univariate
pdf $f_{i}'s$ and joint pdf $f$. The density of the Gaussian copula with the
covariance matrix $\Sigma$, \cite{Song.2000}
\begin{equation*}
c(\underline{u})=|\Sigma|^{-\frac{1}{2}}\exp\bigg\{-\frac{1}{2}\q^T(\Sigma^{-1}-\I_p)\q\bigg\},
\end{equation*}
where $\underline{u}=\{u_1,u_2,\ldots,u_p\}$, $\q=(q_1,\ldots,q_p)^T$, with
$q_j=\Phi^{-1}(u_j)$ for $j=1,2,\ldots,p$ and $\Phi$ is the cdf of $N(0,1)$.
Note that $u_j=F_j(\omega_j)$, could be any distribution. The density of $t$
copula \cite{Demarta.2005} has the form
\begin{equation}\label{eqn_t_copula_density}
c_{\nu}^{t}(\underline{u})=\frac{f_{\nu,\Sigma}(t_{\nu}^{-1}(u_1),\ldots,t_{\nu}^{-1}(u_p))}{\prod_{j=1}^{p}f_{\mu}(t_{\nu}^{-1}(u_i))},~~\underline{u}\in (0,1)^p,
\end{equation}
where $f_{\nu,\Sigma}$ is the joint density of $p$-variate multivariate
$t$-distributions $t_{p}(\nu,0,\Sigma)$ with $\nu$ degrees of freedom,
$\Sigma$ is the covariance matrix and $f_{\nu}$is the standard density of
univariate $t$-distribution with $\nu$ degrees of freedom. The joint prior
density function is, by differentiating (\ref{eqn_Sklar_copula})
\begin{equation}\label{eqn_jnt_copula_prior_pdf}
f(\bfb)=c[F_1(\omega_1),F_2(\omega_2),\ldots,F_p(\omega_p)]\prod_{j=1}^{p}f_j(\omega_j),
\end{equation}
where $c$ is the density of $C$ and $f_1,\ldots,f_p$ are marginal prior
densities. Now we present the LGC prior.

\subsection{Lasso with Gauss-Copula Prior}
\begin{figure*}[ht!]
\centering
\includegraphics[width=0.8\textwidth]{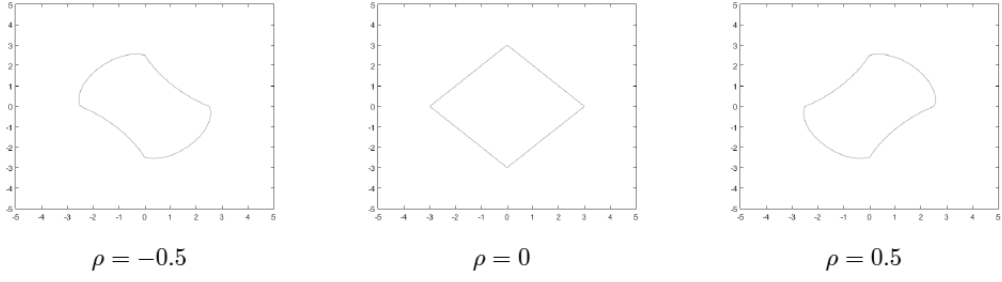}
\caption{Two dimensional contour plot for LGC prior for different values of correlation parameter ($\rho$).
 When $\rho=0$ it represents lasso penalty. For non-zero $\rho$
 contour plot represents non-convex penalty structure. In experiments, $\rho$ is
 learned from data.}
\label{fig_lasso_gauss_copula}
\end{figure*}

Suppose, $F_{L:j}(\omega_j)$ is the marginal prior cdf of the Laplace
distribution over $\omega_j$, $f_{L:j}$ is the marginal prior pdf of Laplace
distribution and consider Gauss copula for $c$ in
(\ref{eqn_jnt_copula_prior_pdf}), then we get the joint prior pdf for $\bfb$
as LGC prior, where
\begin{equation}\label{eqn_lasso_Gauss_copula_density}
\begin{split}
c(\underline{u})=|\Sigma|^{-\frac{1}{2}}\exp\bigg\{-\frac{1}{2}\q^T(\Sigma^{-1}-\I_p)\q\bigg\},
\end{split}
\end{equation}
where $\underline{u}=\{F_{L:1}(\omega_1),\ldots,F_{L:p}(\omega_p)\}$,
$\q=(q_{1},\ldots,q_{p})$ with $q_{j}=\Phi^{-1}(F_{L:j})~\forall j$.
Substituting equation (\ref{eqn_lasso_Gauss_copula_density}) into the
equation (\ref{eqn_jnt_copula_prior_pdf}) would yield the analytical
expression of the joint prior pdf of LGC prior. Assuming
that the density function $f_{L}$ is a Laplace pdf with location parameter 0
and scale parameter as $\lambda$, as in the lasso prior, the final expression
would be
\begin{eqnarray}\label{eqn_jnt_lasso_gauss_copula_prior_pdf}
f(\bfb)&=&|\Sigma|^{-\frac{1}{2}}\exp\bigg\{-
           \frac{\q^{T}(\Sigma^{-1}-\I_{p})\q}{2}\bigg\}
           {\frac{\lambda}{2}}^{p} \exp\big\{-\lambda\sum_{i=1}^{p}|\omega_{i}|\big\}.
\end{eqnarray}
In figure (\ref{fig_lasso_gauss_copula}), we present joint prior pdf of
LGC prior for the two dimensional case (i.e., $\omega_1$ and
$\omega_2$).

\begin{itemize}
\item Note that if we consider $\Sigma=\I_p$, then the simple lasso prior
    becomes special case of the LGC prior
    (\ref{eqn_jnt_lasso_gauss_copula_prior_pdf}). These arguments can be
    seen from the figure (\ref{fig_lasso_gauss_copula}), where the contour
    plots for the LGC prior are shown, for different
    values of $\rho$ (correlation parameter). In practice, $\rho$ is
    learned from data.
\item The advantage of LGC prior is that it can
    include the structural dependence among the predictor variables.
    Due to the sharp edges of LGC, it can do subset
    selection like the lasso or EN.
\item One disadvantage of lasso is that it usually fails to do group
    selection, i.e., it gives inaccurate solutions when features
    are correlated. Similar to the EN, the LGC can
    deal with this problem by introducing correlation and making it as a
    favourable choice for a regularizer.
\item Choice of copula can vary the nature of regularizer and hence
  the final answer. In the experiment, presented in section
  (\ref{section:experiments}), the LTC and LGC yielded different
  solutions and standard error. Thus copula selection is also a
  possibility when we have a large number of choices for copula. For a
  small number of copula choices we can use cross-validation, but for
  a large number of copula choices, there is a need for copula
  selection. However, in this paper, we restrict ourselves only to the
  LGC, LTC and its applications.
\end{itemize}
A desirable supervised learning task for generic data should have the following properties.
\begin{itemize}
\item It should be able to make automatic feature selection.
\item It should work in the case of $p>n$.
\item It should be able to make a group selection for the correlated predictors.
\end{itemize}
Copula prior enjoys all the above qualities. It can make automatic
feature selection, can work for higher dimensions, and can do grouped
selection due to its built-in correlation structure.

\subsection{Lasso with $t$-Copula Prior}

The density of $t$ copula \cite{Demarta.2005} has the form
\begin{equation}\label{eqn_t_copula_density}
c_{\nu}^{t}(\underline{u})=\frac{f_{\nu,\Sigma}(t_{\nu}^{-1}(u_1),\ldots,t_{\nu}^{-1}(u_p))}{\prod_{j=1}^{p}f_{\mu}(t_{\nu}^{-1}(u_j))},~~\underline{u}\in (0,1)^p,
\end{equation}
where $f_{\nu,\Sigma}$ is the joint density of $p$-variate multivariate
$t$-distributions $t_{p}(\nu,0,\Sigma)$ with $\nu$ degrees of freedom,
$\Sigma$ is the covariance matrix and $f_{\nu}$is the standard density of
univariate $t$-distribution with $\nu$ degrees of freedom. The joint prior
density function is, by differentiating (\ref{eqn_Sklar_copula})
\begin{equation}\label{eqn_jnt_copula_prior_pdf}
f(\bfb)=c[F_1(\omega_1),F_2(\omega_2),\ldots,F_p(\omega_p)]\prod_{j=1}^{p}f_j(\omega_j),
\end{equation}
where $c$ is the density of $C$ and $f_1,\ldots,f_p$ are marginal prior
densities. Now we consider the $F_{L:j}(\omega_j)$ as the marginal prior cdf
of the Laplace distribution over $\omega_j$, $f_{L:j}$ is the marginal prior
pdf of Laplace distribution and consider $t$-copula for $c$ in
(\ref{eqn_t_copula_density}), then we get the joint prior pdf for $\bfb$ as
`lasso $t$ copula' (LTC) prior, where
\begin{eqnarray}
\label{eqn_jnt_prior_pdf_t_copula_v1}
&& \log f(\omega_1,\ldots,\omega_p|\Sigma) = \stackrel{part~1}{\overbrace{\log
                                          \big(c(F_{L:1}(\omega_1),\ldots,F_{L:p}(\omega_p))\big)}}+\stackrel{part~2}{\overbrace{\log\big(\prod_{j=1(1)p}f_{L:j}(\omega_j)\big)}}.
\end{eqnarray}
In (\ref{eqn_jnt_prior_pdf_t_copula_v1}), after some simplification, the part
1 can be expressed as,
\begin{equation}
\begin{split}
&\log\big(c(F_{L:1}(\omega_1),\ldots,F_{L:p}(\omega_p))\big)=p\log\Bigg(\frac{\Gamma(\frac{\nu+1}{2})}{\Gamma(\frac{\nu}{2})}\Bigg)\\
&-\bigg(\frac{\nu+p}{2}\bigg)\log\bigg(1+\frac{\q^T\Sigma^{-1}\q}{\nu}\bigg)
+\log \bigg(\frac{\Gamma(\frac{\nu+p}{2})}{\Gamma(\frac{\nu}{2})|\Sigma|^{1/2}}\bigg)\\
&+\sum_{j=1(1)p}\bigg(\frac{\nu+1}{2}\bigg)\log\bigg(1+\frac{q_j^2}{\nu}\bigg)
\end{split}
\end{equation}
where $\q=(q_{1},\ldots,q_{p})$ with $q_{j}=t_{\nu}^{-1}(F_{L:j})~\forall j$.
The part 2 of (\ref{eqn_jnt_prior_pdf_t_copula_v1}) can be expressed as
\begin{eqnarray*}
\log\big(\prod_{j=1(1)p}f_{L:j}(\omega_j)\big)&=&\log\bigg[\prod_{j=1(1)p}\frac{\lambda}{2}\exp\big(-\lambda|\omega_j|\big)\bigg],\\
&=&p\log\big(\frac{\lambda}{2}\big)-\lambda \sum_{j=1}^p|\omega_j|.
\end{eqnarray*}
Hence the joint prior density in log-scale for LTC prior can be
expressed as
\begin{eqnarray*}
&&\log f(\omega_1,\ldots,\omega_p)
   =-\bigg(\frac{\nu+p}{2}\bigg)\log\bigg(1+\frac{\q^T\Sigma^{-1}\q}{\nu}\bigg)\\
&&~~~+\sum_{j=1(1)p}\bigg(\frac{\nu+1}{2}\bigg)\log\bigg(1+\frac{q_j^2}{\nu}\bigg)\\
&&~~~+\log
   \bigg(\frac{\Gamma(\frac{\nu+p}{2})}{\Gamma(\frac{\nu}{2})|\Sigma|^{1/2}}\bigg)+p\log\Bigg(\frac{\Gamma(\frac{\nu+1}{2})}{\Gamma(\frac{\nu}{2})}\Bigg)\\
&&~~~+p\log\big(\frac{\lambda}{2}\big)-\lambda \sum_{j=1}^p|\omega_j|.
\end{eqnarray*}

\begin{figure*}[ht]
\centering
\includegraphics[width=0.80\textwidth]{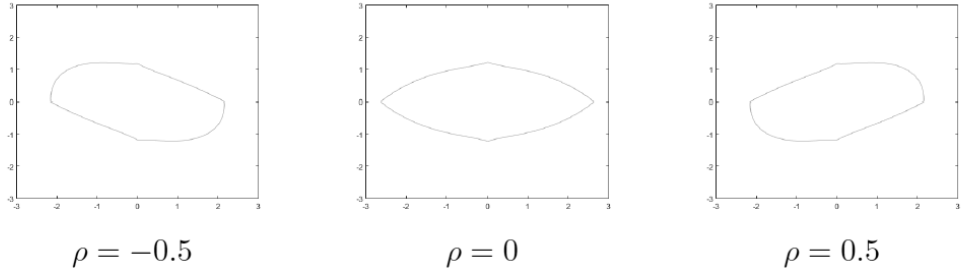}
\caption{Two dimensional contour plot for lasso-$t$-copula prior with $\nu=10$ for different values of correlation parameter ($\rho$). For $\rho=0$ the shape deflects from lasso.}
\label{fig_lasso_t_copula}
\end{figure*}

\begin{figure*}[ht]
\centering
\includegraphics[width=0.80\textwidth]{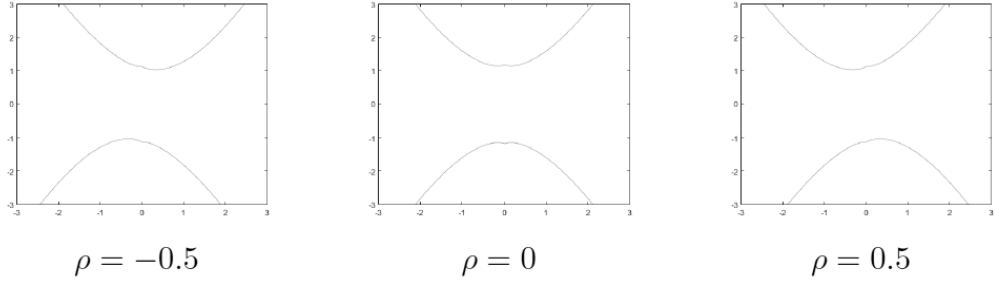}
\caption{Two dimensional contour plot for lasso-$t$-copula prior with $\nu=1$ for different values of correlation parameter ($\rho$). Also known as Cauchy copula.}
\label{fig_lasso_t1_copula}
\end{figure*}

If we consider $\Sigma=\I$, then
$\log\big(c(F_{L:1}(\omega_1),\ldots,F_{L:p}(\omega_p))\big)$
would still be non-zero. Hence, unlike LGC prior, with zero correlation
    among the coefficients, the shape of the $t$-copula prior deflect from
    lasso prior. The argument can be seen clearly from figure
    (\ref{fig_lasso_t_copula}), where the contour plots for the
    LTC prior (with $\nu=10$) are shown, for different values
    of correlation parameter.

In figure (\ref{fig_lasso_t1_copula}) we present the contour plot of
    lasso with $t$-copula with the degrees of freedom to be $\nu=1$. This is
    essentially Cauchy copula. Apparently, the contour plot for Cauchy copula
    shows a very undesirable property.

\subsection{Optimization}
The standard approach would be to develop the full Bayesian solution to
estimate the posterior mean of $\bfb$ via MCMC technique
\cite{Park.2008,Li.Lin.2010,Kyung.2010}. However, we have to prove the
geometric ergodicity of the Markov chains \cite{Khare.Hobert.2013,
  Roy.2017} for our proposed copula prior. This would be a significant
detour from the current paper. Hence we set aside this work for another article for which we are currently working on.

In this paper, we implement the posterior mode of $\bfb$. Please note that
posterior mode is Bayes estimator under Kullback-Leibler type loss function
\cite{Das.Dey.2010}. We estimate the posterior mode via augmented Lagrangian
optimization technique \cite{Conn.1991,
  Birgin.2008}. This method
consolidates the objective function and the nonlinear penalty into a single
function. Here the objective function is the negative log of the likelihood
function, and the `penalty' is the negative log of copula prior. The
mathematical form of the objective function with copula regularizer would be
as follows,
\begin{equation*}
L(\bfb)=\|\y-\x\bfb\|_{2}^{2}-\ln f(\omega_1,\ldots,\omega_p|\Sigma,\lambda).
\end{equation*}
Now using (\ref{eqn_jnt_copula_prior_pdf}), and as marginals are
from Laplace distribution, we can write above equation as follows,
\begin{eqnarray}
\label{eqn_log_Post}
L(\bfb)&=&\|\y-\x\bfb\|_{2}^{2}-\ln
           \big(c(F_{L:1}(\omega_1),\ldots,F_{L:p}(\omega_p))\big)
           +\lambda \sum_{j=1}^p|\omega_j|.
\end{eqnarray}

Above equation is an unconstrained minimization problem. However, we
cannot use the augmented Lagrangian algorithm here, because it
requires the objective function and constraints to be twice
continuously differentiable. The presence of $\ell_{1}$ norm of $\bfb$ vector makes it not
differentiable at 0. Since $|\omega_j|$ is not differentiable at 0 for all
$j$, we do transformation to make it a continuous function. The approach is to split the
element $\omega_{j}$ of vector $\bfb$ into $\omega_{j}^{+}$ and
$\omega_{j}^{-}$ so that $\omega_{j}=\omega_{j}^{+}-\omega_{j}^{-}$. If
$\omega_{j}>0$, then we have $\omega_{j}^{+}=|\omega_{j}|$ and
$\omega_{j}^{-}=0$, else we will have $\omega_{j}^{-}=|\omega_{j}|$ and
$\omega_{j}^{+}=0$. Mathematically we can write
$$
\omega_{j}^{+}=\frac{|\omega_{j}|+\omega_{j}}{2},~~ and~~ \omega_{j}^{-}=\frac{|\omega_{j}|-\omega_{j}}{2}.
$$
Both $\omega_{j}^{+}$ and $\omega_{j}^{-}$ are non negative numbers. Main
advantage of this splitting is that now we can express
$|\omega_{j}|=\omega_{j}^{+}+\omega_{j}^{-}$, hence effectively we can now
avoid the absolute values. Substitute
$\omega_{j}=\omega_{j}^{+}-\omega_{j}^{-}$ and
$|\omega_{j}|=\omega_{j}^{+}+\omega_{j}^{-}$ into (\ref{eqn_log_Post}) we get
the final non linear optimization problem,
\begin{eqnarray}\label{eqn_obj}
\min_{\bfb^{+},\bfb^{-}}& ~& L(\bfb^{+},\bfb^{-}),\nonumber \\
\text{subject~to}        &~&
\sum_{j=1}^p\omega_j^{+}\omega_{j}^{-}=0, \nonumber \\
\omega_{j}^{+},\omega_{j}^{-}&\geq&  0 ~~~ \forall j.
\end{eqnarray}
The objective function and constraints in (\ref{eqn_obj}) both are continuous functions.
Now we can use the augmented Lagrangian optimization on (\ref{eqn_obj}), find the optimal $\bfb^{+}$ and $\bfb^{-}$
and estimate the final solution as $\bfb^{*}=\bfb^{+}-\bfb^{-}$.

We used the analytical gradient of LGC prior, to
speedup the optimization procedure for big data with large number of
features. 

\subsection{Tuning of Hyperparameters}
The two unknown parameters for the LGC and LTC are the scale parameter
$\lambda$ and the variance-covariance matrix $\Sigma$. The dimensionality of
data plays a significant role in the estimation of $\Sigma$. For $n>p$ case
we can determine the prior correlation between the coefficients using the
covariance of predictors, i.e., $(X^{T}X)$. However for $n<p$ case, we cannot
use the covariance matrix, though it preserves the variance-covariance
structure of the data. Choosing $\Sigma$ as identity matrix in $n<p$ case
could be a poor choice if the features are highly correlated. Hence the Ridge
prior seems to be a compromise between the actual data covariance and the
Identity matrix. Following this idea we choose $\Sigma$ to be as follows
\begin{eqnarray}\label{Sigma}
\Sigma =\frac{(X^{T}X+cI)}{(1+c)},     ~~ if ~ n < p.
\end{eqnarray}
and
\begin{eqnarray*}
\Sigma =(X^{T}X),     ~~ if ~ n \geq p.
\end{eqnarray*}
Here $p$ is the number of features and $c$ is a constant. To maintain the
variance-covariance structure of the data we would usually choose a very
small value of c. The scale parameter $\lambda$ is the other parameter, which
we would like to learn. For a given correlation as $\lambda$ increases the
copula penalty function also increases. We estimate the scale parameter
$\lambda$ via $10$-fold cross validation technique
\cite{Hastie.Tibshirani.Friedman.2008}.

\subsection{Analytical Gradient of LGC prior}
We used the analytical gradient of LGC prior, to speedup the optimization
procedure for big data with large number of features. The expression for the
gradient of squared loss function with LGC regularizer
L$(\omega^{+},\omega^{-})$, is as follows:
\begin{eqnarray}
\frac{\partial L}{\partial \omega^{+}}=-2X^{T}(y-X\omega)+(\Sigma^{-1}-\I)K^{+}\q+\lambda\text{\textbf{1}}, \nonumber\\
\frac{\partial L}{\partial \omega^{-}}=2X^{T}(y-X\omega)+(\Sigma^{-1}-\I)K^{-}\q+\lambda\text{\textbf{1}}.\nonumber,
\end{eqnarray}
where $\omega=\omega^{+}-\omega^{-}$, $K^{+}$ and $K^{-}$ are diagonal
matrices with
$K^{+}(i,i)=\frac{dq(\omega_{i})}{d\omega_{i}}\text{sgn}(\omega_{i}^{+})$,
and $K^{+}(i,j)=0\quad \forall~ i\neq j$. Similarly we can express the matrix
$K^{-}$ as
$K^{-}(i,i)=\frac{dq(\omega_{i})}{d\omega_{i}}\text{sgn}(\omega_{i}^{-})$,
and $K^{-}(i,j)=0\quad \forall ~i\neq j$. The $sgn(.)$ represents the signum
function.

\subsection{Archimedean Copula}

Other than elliptical copula (like Gauss or $t$ copula), the Archimedean copula provides the big class of models. For example Clayton copula \cite{Clayton.1978}, Frank copula\cite{Frank.1979} or  Gumbel copula \cite{Gumbel.1960} are popular Archimedean copulas. However, it's worth noting that the Archimedean copulas with the dimension three or higher only allows positive association. The bivariate Archimedean copulas can handle the negative association. This undesirable feature of the Archimedean copulas makes it an unlikely candidate to be considered as the copula prior for $\bfb$.

\section{Results on Copula Prior}
In this section we present some important theoretical results for
LGC prior. The pdf of LGC prior can be expressed as
(\ref{eqn_jnt_lasso_gauss_copula_prior_pdf}). The nature of $q_j$ in (\ref{eqn_jnt_lasso_gauss_copula_prior_pdf}) is crucial in determining the nature of
the LGC regularizer, where $q_j$ is monotonic function of
$\omega_j$. Here we consider the following assumptions. In the
appendix \ref{Appendix_A}, we present graphical support for the
assumptions in figure (\ref{assumption_1}) and (\ref{assumption_4}).

\begin{assumption}\label{A1}
When $\omega_{j}>0$, then $q_{j}$ is concave in nature and for
$\omega_{j}<0$, $q_{j}$ is a convex function.
\end{assumption}

\begin{assumption}\label{A4}
If $\hat{\omega}_{k}\neq\hat{\omega}_{l}$ have the same sign then the
following inequality holds.
\begin{eqnarray*}\label{12}
&&\frac{q(\hat{\omega}_{k}+\hat{\omega}_{l})[q(\hat{\omega}_{k})\frac{dq(\hat{\omega}_{l})}{d\omega}
-q(\hat{\omega}_{l})\frac{dq(\hat{\omega}_{k})}{d\omega}]}{\frac{dq(\hat{\omega}_{l})}{d\omega}-\frac{dq(\hat{\omega}_{k})}{d\omega}
}\nonumber \\
&<&\frac{q(\hat{\omega}_{k})^{2}\frac{dq(\hat{\omega}_{l})}{d\omega}
-q(\hat{\omega}_{l})^{2}\frac{dq(\hat{\omega}_{k})}{d\omega}}{\frac{dq(\hat{\omega}_{l})}{d\omega}-\frac{dq(\hat{\omega}_{k})}{d\omega}}.
\end{eqnarray*}
\end{assumption}

\begin{lemma}\label{lma_1}
An unique solution for LGC penalty always exists.
\end{lemma}

\begin{lemma}\label{lma_2}
Under the assumption that each predictor is standardized, and $\Sigma = X^TX$
in (\ref{eqn_jnt_lasso_gauss_copula_prior_pdf}) then joint prior pdf of LGC
in (\ref{eqn_jnt_lasso_gauss_copula_prior_pdf}) can be expressed as
\begin{equation}\label{eqn_jt_pdf_LGC_approx_form_final}
f(\bfb)=-\sum_{i=1}^{p}q_{i}\sum_{j\neq i}q_{j}\rho_{ji}^{*}+
\lambda \sum_{j=1}^{p}|\omega_{j}|
\end{equation}
where $\rho_{ij}^{*}$ is partial correlation. Note that $\Sigma^{-1}$ the
covariance matrix, is converted into a correlation matrix.
\end{lemma}

\begin{lemma}\label{lma_3}
For $k,l \in \{1,2,\hdots,p\}$, if $x_{k}\approx x_{l}$ then  $\rho_{jk}^*\approx
\rho_{jl}^* ~~~\forall j\neq \{k,l\}$
\end{lemma}

\begin{lemma}\label{lma_4}
 If $x_{k}\approx x_{l}$ and $\rho_{kl}^{*}\textgreater 0$ $
\Rightarrow \hat{\omega}_{k}\approx\hat{\omega}_{l}$.
\end{lemma}

\begin{theorem}\label{thm_LGC_group_effect}
Given data $\y$, $\X$ and parameters $\lambda$, the response $\y$ is
centred and the predictors $\X$ are standardized. Let $\hat{\omega}_{k},
\hat{\omega}_{l}$ be the LGC estimate. Suppose that
$\hat{\omega}_{k}\geq\hat{\omega}_{l}>0$.
Define
\begin{equation*}
D_{\lambda}(k,l)=|q_{k}-q_{l}|\frac{dq_{k}}{d\omega_{k}}
\end{equation*}
then
\begin{eqnarray}
&&D_{\lambda}(k,l)\leq
  |y|\frac{\sqrt{2(1-\rho_{lk})}}{|\rho_{kl}^{*}|} \nonumber \\
&&~~~~+\lambda\frac{\sqrt{\sum_{j\neq
k,l}q_{j}^2\frac{\pi}{2}}\sqrt{\sum_{j\neq
k,l}(\rho_{jl}^{*}-\rho_{jk}^{*})^{2}}}{|\rho_{kl}^{*}|} \label{Ineq}
\end{eqnarray}
\end{theorem}

The unitless quantity $D_{\lambda}(k,l)$ describes the difference between the coefficient paths of
predictors $k$ and $l$. If $x_k$ and $x_l$ are highly correlated, then
theorem (\ref{thm_LGC_group_effect}) says that the difference between
the coefficient paths of predictor $x_k$ and $x_l$ is almost 0. The
upper bound in the inequality in theorem (\ref{thm_LGC_group_effect}) provides a quantitative description for the
grouping effect of the LGC prior.

\section{Learning Copula Prior from Big Data}

In order to handle the `big data,' we present a resample technique for learning
of $\bfb$ with LC prior. We consider training dataset consisting of $n$ independent and
identically distributed samples
$\mathcal{D}_n=\{x_{i},y_{i}\}_{i=1}^{n}$, where $n$ is large. We draw a random re-sample of
subset $\mathcal{D}_m$ of size $m(<n)$ from $\mathcal{D}_n$. We learn $\hat{\bfb}$ from $\mathcal{D}_{m}$ using
   (\ref{eqn_obj}) and repeat the process $M$ times, where $M$ is the
   simulation size. So for the coefficients of each predictor we have
   $M$ solutions and we estimate the final solution by taking the
   median of the $M$ solutions. The algorithm is presented in (1).
\begin{figure}[h!]
\includegraphics[width=17cm]{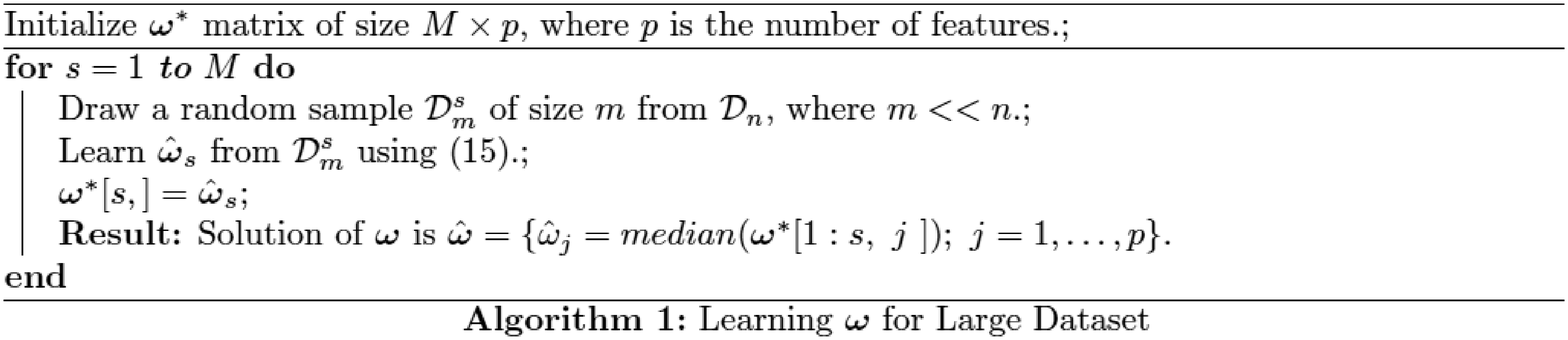}
\label{alg_large_data}
\end{figure}

\section{Experiments}\label{section:experiments}
In this section we implement LC prior on simulated data and real
life examples. We compared the performance of LC prior with Lasso
\cite{Tibshirani.1996} and EN \cite{Zou.Hastie.2005}. For
implementation of EN and Lasso we have used publicly available
packages.

\begin{figure*}[ht]
\centering
\includegraphics[width=16cm, height=4cm]{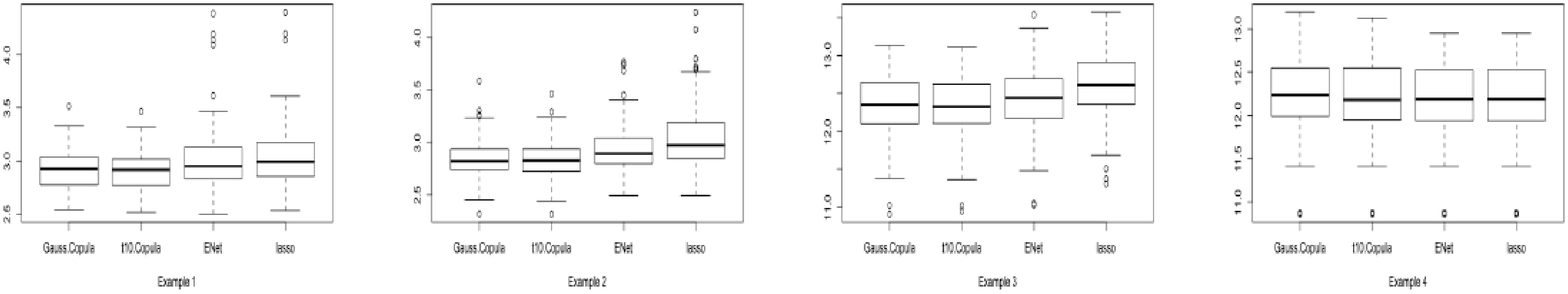}
\caption{Comparing the RMSE of the LGC, the LTC, the EN and the lasso. The LGC and t-copula outperform the EN and lasso in examples 1, 2 and 3. In example 4, the performance among the methods is similar. It indicates that copula prior often perform well in the presence of high correlation between the predictors.}
\label{fig_simulation_boxplot}
\end{figure*}

\begin{table*}[ht]
\centering
\begin{tabular}{l|rrrr}
  \hline
 & Example 1 & Example 2 & Example 3 & Example 4 \\
  \hline
  LGC     & 2.92 (0.02) & 2.82 (0.020) & 12.35 (0.051)& 12.24 (0.07)\\
  lasso $t_{\nu=10}$ Copula & 2.92 (0.02) & 2.83 (0.017) & 12.32 (0.048)& 12.18 (0.08)\\
  Elastic net & 2.95 (0.03) & 2.89 (0.025) & 12.44 (0.068)& 12.19 (0.07) \\
  lasso & 2.99 (0.02) & 2.97 (0.026) & 12.61 (0.043)& 12.19 (0.07)\\
   \hline
\end{tabular}
\caption{Median RMSE for the simulated examples and
four methods based on 100 replications. The numbers in parentheses are the corresponding standard errors (of the medians) estimated by using the bootstrap with $B=1000$ resamplings on the 100 RMSE.}
\label{table_sumulation}
\end{table*}

\begin{figure*}[ht]
\centering
\includegraphics[width=0.60\textwidth]{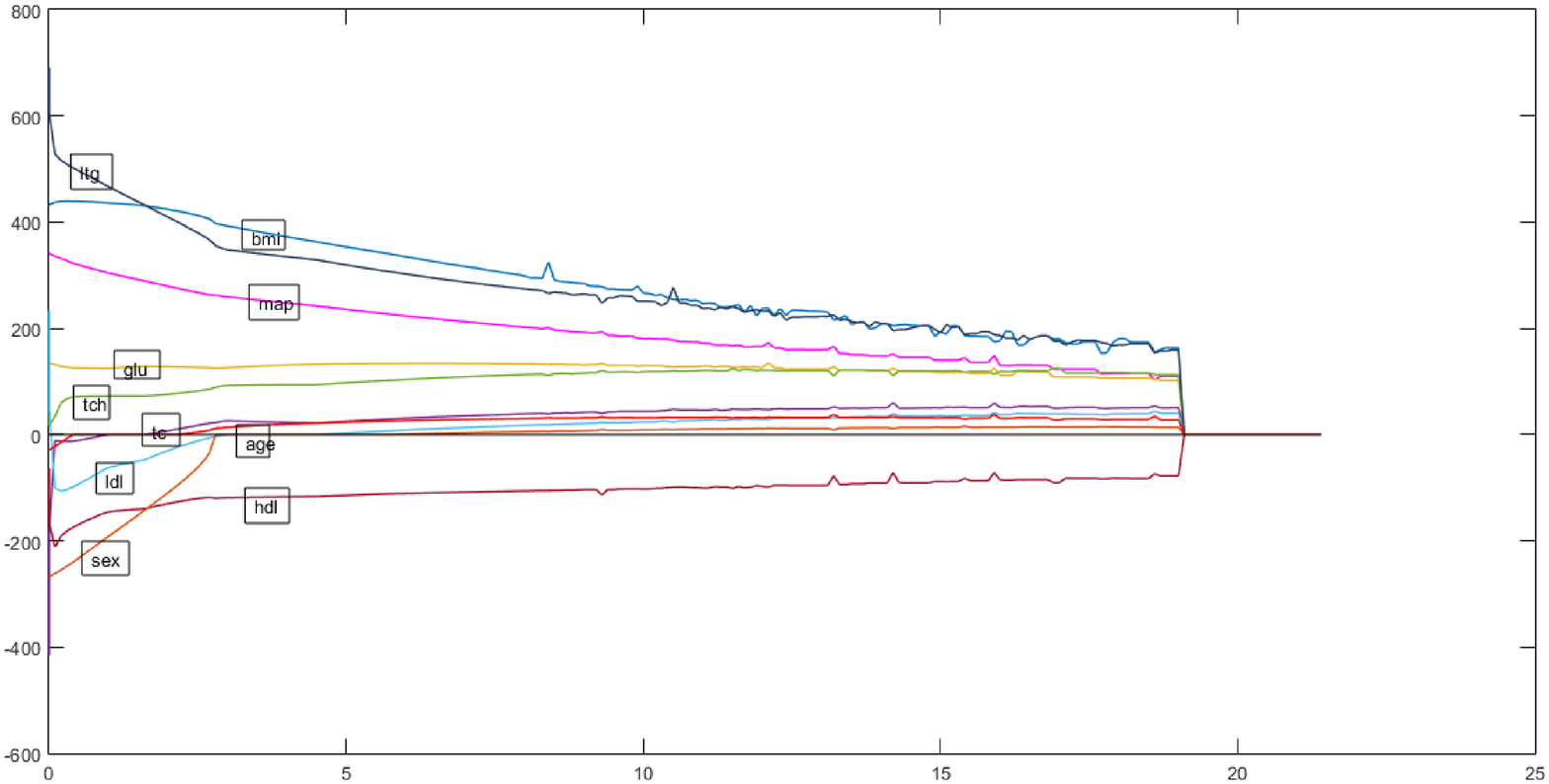}
\caption{Solution Path of the LGC in Diabetes Datset}
\label{fig_diabetes_Gaussian_Copula_Path}
\end{figure*}

\begin{table*}[ht]
\centering
\begin{tabular}{l|rrr}
  \hline
 & Test set MSE & Tuning Parameter Estimates & Selected Features  \\
  \hline
Lasso & 897.071 & $\lambda$ = 5.59 & 6/10 \\
Elastic net & 871.368 & $\alpha$ = 0.947, $\lambda$ = 2.915 & 6/10 \\
Gaussian Copula  & 856.693 & $\lambda$ =0.458 & 10/10 \\
t Copula ($\nu=10$)  & 857.576 & $\lambda$ = 0.615 & 10/10 \\
   \hline
\end{tabular}
\caption{Out-Sample Mean-Square Error (MSE) in Diabetes Dataset}
\label{tabl_diabetes_data_mse}
\end{table*}

\begin{table*}[ht]
\centering
\begin{tabular}{l|rrr}
  \hline
 & Test set MCE & Tuning Parameter Estimates & Selected Features  \\
  \hline
Lasso & 9/22 & $\lambda$ =0.09 & 7/201 \\
Elastic net & 5/22 & $\alpha$ =0 , $\lambda$ =0.03 & 201/201  \\
Gaussian Copula  & 5/22 & $\lambda$ = 0.1 & 70/201 \\
   \hline
\end{tabular}
\caption{Out-Sample Misclassification Error (MCE) in Colon Cancer Dataset}
\label{tabl_Alon_data_mce}
\end{table*}

\begin{table*}[ht]
\centering
\begin{tabular}{l|rrr}
  \hline
 & Test set MSE & Tuning Parameter Estimates & Included Features  \\
  \hline
Lasso & 0.699 & $\lambda$ = 0.01 & 16/27 \\
Elastic net & 0.699 & $\alpha$ = 0.21, $\lambda$ = 0.03 & 18/27 \\
Gaussian Copula  & 0.768 & $\lambda$ = 0.14 & 4/27 \\
   \hline
\end{tabular}
\caption{Out-Sample Mean Square Error (MSE) in Energy Dataset}
\label{tabl_energy_data_mse}
\end{table*}

\subsection{Synthetic experiments}
Here we present four examples from \cite{Tibshirani.1996,Zou.Hastie.2005},
to compare the prediction performance of the lasso and EN and
proposed copula prior. For each example, our simulated data consist of a
training data set, an independent validation data set, and a separate test
data set. The validation data sets were used to select the tuning parameters
and then the models were fitted on the training data set. We computed the
test error (the mean-squared error) on the test data set. The simulation
scenarios are as follows:

\begin{itemize}
\item[(a)] We consider the true model, $\y=\X\bfb+ \epsilon, ~~\epsilon\sim
    N(0,\sigma^2\I)$, where we set $\bfb=(3, 1.5, 0, 0, 2, 0, 0, 0)$,
    $\sigma=3$ and $\X_k\sim MVN_p(0,\Sigma),~k=1,\hdots,p$, where
$$
\Sigma=(\sigma_{ij})=\bigg\{\begin{array}{cc}
1 & i=j,\\
0.95 & i\neq j.
\end{array}
$$
We simulated 100 data sets. Training set sample size = 20; Validation set
sample size = 20; and Test set sample size = 200.
\item[(b)] Example 2 is the same as example 1, except that $\omega_j=0.85$
    for all $j$.
\item[(c)] In example 3, we consider the true model, $\y=\X\bfb+ \epsilon,
    ~~\epsilon\sim N(0,\sigma^2\I)$, where we set
$$
\bfb=(\underbrace{0,\hdots,0}_{10},\underbrace{2,\hdots,2}_{10},\underbrace{0,\hdots,0}_{10},\underbrace{2,\hdots,2}_{10}),
$$
and $\sigma=15$ and $\X_k\sim MVN_p(0,\Sigma),~k=1,\hdots,p$, where
$$
\Sigma=(\sigma_{ij})=\bigg\{\begin{array}{cc}
1 & i=j,\\
0.95 & i\neq j.
\end{array}
$$
We simulated 100 data sets with Training set sample size = 100; Validation
set sample size = 100; and Test set sample size = 400.
\item[(d)] In example 4, we consider the true model, $\y=\X\bfb+ \epsilon,
    ~~\epsilon\sim N(0,\sigma^2\I)$, where we set
$$
\bfb=(\underbrace{3,\hdots,3}_{15},\underbrace{0,\hdots,0}_{25}),
$$
and $\sigma=15$. Let $\epsilon_i\stackrel{iid}{\sim} N(0,0.16)$, for
$i=1,2\hdots,15$, the predictors are generated from
\begin{eqnarray*}
X_i&=&Z_1+\epsilon_i, ~~~Z_1\sim N(0,1),~~~i=1,\hdots,5,\\
X_i&=&Z_2+\epsilon_i, ~~~Z_2\sim N(0,1),~~~i=6,\hdots,10,\\
X_i&=&Z_3+\epsilon_i, ~~~Z_3\sim N(0,1),~~~i=11,\hdots,15.
\end{eqnarray*}
We simulated 100 data sets with Training set sample size = 100; Validation
set sample size = 100; and Test set sample size = 400.
\end{itemize}
Table \ref{table_sumulation} and figure \ref{fig_simulation_boxplot} (box
plots) summarize the prediction results. We see that in the examples 1, 2 and
3, based on RMSE, the `lasso with $t_{\nu=10}$-copula' and LGC tend to
be more accurate than the lasso and the EN. While in example 4, the copula priors tend to do as well as the EN and lasso.
This is expected because in example 4, predictors are not correlated and
hence proposed copula priors will do as good as the regular lasso or the
EN penalty.

\subsection{Regression for Diabetes Data}\label{section_diabetes_dataset} The diabetes dataset arises from the study of 442 diabetes patients describes in the \cite{Efron.2004}. It comprises a sample of 442 diabetic patients. The independent variables are age, sex, body mass index (BMI), average blood pressure, and six blood serum measurements. The dependent variable is the quantitative measure of disease progression in one year after measuring the independent variables. For data analysis, we randomly split the data with 300 observations in the training set and the remaining 142 observations in the test set. We evaluate the MSE for copula prior, EN, and lasso
on the test data set. The table (\ref{tabl_diabetes_data_mse})
presents the results.

For choosing the optimal tuning parameters, we have used ten-fold
cross-validation for all the above regularizers. The LGC prior selects all
the variables in the final model. The unique solution path of LGC prior
(presented in figure \ref{fig_diabetes_Gaussian_Copula_Path}) is formed as it
takes into account the correlation among the predictors. The regularization
paths for lasso and EN for the diabetes dataset are reported in
\cite{Vidaurre.2013}. The six blood serum measurement variables (TC, LDL,
HDL, TCH, LTG, GLU)  are highly correlated with each other since they belong
to the same person's blood. The BMI and map variable also have a significant
amount of correlation with other predictors. For all the three cases the
regularization path for age and sex variable is similar, as the age and sex
variable are not correlated with other predictors. However, the solution path
of LGC prior changes concerning lasso prior, as the LGC prior takes into
account the correlation among the predictors. As a result, it results in
lower MSE than regular lasso and EN.

Both the EN and lasso select sex, BMI, MAP, HDL, LTG, and GLU to be the
significant predictors. EN performs better than lasso due to the presence of
ridge penalty but tends to perform worse than the copula prior. The lasso is
a particular case of the LGC prior when the prior correlation among the
predictors is assumed 0. We have used $t$ copula with 10 degrees of freedom.
The optimal $\lambda$ value in case of $t$ copula comes out to be 0.615.
Again in the copula prior, the test data error is small, as compared with EN and
lasso; $t$ copula selects all the variables in the final model.

\subsection{Classification for Colon Cancer Data}
Microarray data is a classic example of high dimensional data. The
experiments on DNA, RNA and protein microarrays which consists of the
expression state of a vast number of genes generates high dimensional
data. There are often thousands of features (gene expression) for such
data but very few samples. As a result, there is a need for feature
selection in such type of data. The response variable often classified
as the cancerous cell or healthy cell.

Here, we consider the example of the Colon cancer data set as explained in
\cite{Alon.1999}. This dataset consists of 62 tissue samples collected from
colon cancer patients. From these 62 samples, 40 biopsies are from tumors
(labeled as ``1'') and rest 22 biopsies (labeled as ``0'') are from healthy
parts of the colon of the same patients. Top 2000 features are selected from
6500 features based on the confidence in measured gene expression levels.

The goal is to develop a diagnostic rule based on the gene expression
of 2000 genes to differentiate cancerous tissues from healthy
tissues. For this classification problem, we fit a logistic regression
on training data with LGC prior, as the regularizer function. After
learning the coefficients on training data, we use these coefficients
on test data to evaluate the misclassification error.

We divide the data into test and training data set. In the training dataset,
there are 40 tissue samples of which 13 are the normal tissues, and rest 27
are tumor samples.  The misclassification error is evaluated on the remaining
22 samples.  Since the data has very high dimension, we first select top 200
predictors based on their $t$ statistic scores from training data. It helps
us in making the computation easier.

The prior correlation matrix between coefficients learned from
(\ref{Sigma}). Another unknown quantity is the scale parameter
$\lambda$, which we learned from five-fold cross-validation
method. Table (\ref{tabl_Alon_data_mce}) compares the
out-sample misclassification error (MCE) of the LGC prior with other feature selection
methods like lasso $\&$ EN. The LGC prior has much lower
MCE than lasso. Both LGC and EN have the same
MCE, but LGC resulted in a sparse
representation. The LGC selected only 70 features out of 201 features,
whereas the EN selected all the features.

\subsection{Large Time Series Data for Energy and Housing}
The energy appliances dataset arises from the study of household
energy uses from appliances describes in the
\cite{Candanedo.2017}. The dataset is available for about 4.5 months
at 10 minutes interval. The house temperature and humidity conditions
were monitored with a wireless sensor network. The energy data was
logged every 10 minutes. Weather data collected from the nearest
airport weather station (Chievres Airport, Belgium) was downloaded
from a public dataset and merged with the experimental datasets using
the date and time column. Overall the data is a time series having
19735 observations and 27 features. We considered the first 3.15 month
($\approx 70\%$) as training data set, and the rest of dataset as the
test dataset for measuring the performance of the model.

Since the data is a time series, we checked the stationarity of
the variables involved. The Augmented Dickey-Fuller test confirms that
all the variables involved are stationary. Hence we used regular
linear time series model for feature selection.

As the size of data is large, we used the resampling technique defined in the
algorithm  (\ref{alg_large_data}). A small resample of size 1500 sampled with
replacement from the training dataset and solutions obtained using
(\ref{eqn_obj}). The process repeated for 200 times. As a result, we have 200
solutions. Finally, the median of these 200 solutions is considered as the
final solution. For choosing the optimal tuning parameters, we used ten-fold
cross-validation for all the regularizers. We evaluated the MSE for LGC, EN,
and lasso on the test data set. The results presented in table
(\ref{tabl_energy_data_mse}). As evident from the table
(\ref{tabl_energy_data_mse}), the MSE for lasso,EN and LGC are similar. This
is expected because in this data set predictors are only slightly correlated.

\section{Conclusion}\label{Conclusion}

We presented the copula prior, a shrinkage and feature selection
method. The LC prior produces a sparse model with good prediction
accuracy while preserving the grouping effect. The empirical results
and simulations demonstrate the better performance of the LC prior and
its superiority over the EN and the lasso. When used in the binary
classification method, the LC prior appears to perform well on
microarray data regarding the misclassification error, and it makes
automatic gene selection.

The LC prior is implemented in standard supervised learning task, like
regression and classification. The copula prior is a generalization of
the EN and lasso, which has been shown to be an essential device for
model fitting and feature selection.  Our method offers other insights
into the lasso and EN, and ways to improve it.

\section*{Appendix A: Proof of Results on Copula Prior}\label{Appendix_A}



\noindent \textbf{Proof of Lemma \ref{lma_3}}: The objective function is
\begin{eqnarray}\label{11_1}
L(\omega_{k},\omega_{l})&=& |\y-\x\bfb |_{2}^{2}-2q_{k}\sum_{j\neq k,j\neq l}\rho_{jk}^{*}q_{j}\nonumber \\
&&~~~~-2q_{l}\sum_{j\neq k,j\neq l}\rho_{jl}^{*}q_{j}-2\rho_{kl}^{*}q_{k}q_{l}+\lambda(|\omega_{k}|+|\omega_{l}|)
\end{eqnarray}
From \cite{Kwan.2014} we know that partial correlation satisfy the following
relation, $\rho_{jk}^{*}=\frac{\hat{\beta}_{jk}}{(1-R_{k}^2)}$, where
$\hat{\beta}_{jk}$ is the ols coefficient of the following regression
equation $x_{k}=\sum_{j\neq k}x_{j}\beta_{jk}+\epsilon_{k}$, and $R_{k}^2$ is
the R square value for this regression equation. By similar argument the partial correlation
$\rho_{jl}^{*}=\frac{\hat{\beta}_{jl}}{(1-R_{l}^2)}$ where $\hat{\beta}_{jl}$
is the ols coefficient of the following regression equation
$x_{l}=\sum_{j\neq l}x_{j}\beta_{jl}+\epsilon_{l}$, and $R_{l}^2$ is the R
square value for this regression equation. The ols coefficients $\hat{\beta_{k}}, \hat{\beta_{l}}$ satisfy the following
linear equation.
\begin{eqnarray}
  \sum_{j\neq k}x_{j}\hat{\beta_{jk}} &=& x_{k}  \label{ols_1} \\
  \sum_{j\neq l}x_{j}\hat{\beta_{jl}}&=& x_{l}      \label{ols_2}
\end{eqnarray}

Subtract (\ref{ols_2}) from (\ref{ols_1}), and using the approximation that
$x_{k}\approx x_{l}$ we will get the following equation.

\begin{equation}
\sum_{j\neq k,l}x_{j}\delta_{j}=0, \label{ols_3}
\end{equation}

where $\delta_{j}=(\hat{\beta_{jk}}-\hat{\beta_{jl}}) \quad \forall$ j $\neq$
k,l. Equation (\ref{ols_3}) is satisfied only if
$\hat{\beta_{jk}}=\hat{\beta_{jl}} \quad \forall$ j $\neq$ k,l. Similarly we
can show that $R_{k}^2$ approaches $R_{l}^2$ as $x_{k}$ approaches $x_{l}$.
Consequently if $x_{k}\approx x_{l}$ then ($\rho_{jk}^{*}\approx
\rho_{jl}^{*})\quad \forall j\neq \{k,l\}$. Q.E.D.

\vspace{2cm}

\noindent \textbf{Proof of Lemma \ref{lma_4}}: Suppose $\hat{\bfb}$ is the optimal solution with
$\hat{\omega}_{k},\hat{\omega}_{l}\textgreater 0$. At the optimal point,
\scalebox{1.3}{$\frac{\partial L}{\partial \omega_{k}}=0$} and
\scalebox{1.3}{$\frac{\partial L}{\partial \omega_{l}}=0$}, so we have
\begin{eqnarray}
-2x_{k}^{T}(\y&-&\x\hat{\bfb})-2\frac{dq(\hat{\omega_{k}})}{\hat{\omega_{k}}}\sum_{j\neq k,j\neq l}
\rho_{jk}^{*}q(\hat{\omega_{j}}) \nonumber\\
&&-2\rho_{lk}^{*}q(\hat{\omega_{l}})\frac{dq(\hat{\omega_{k}})}{\omega_{k}}+\lambda=0 \label{13_1}\\
-2x_{l}^{T}(\y&-&\x\hat{\bfb})-2\frac{dq(\hat{\omega_{l}})}{\omega_{l}}\sum_{j\neq
   k,j\neq l}\rho_{jl}^{*}q(\hat{\omega_{j}})\nonumber \\
&&-2\rho_{lk}^{*}q(\hat{\omega_{k}})\frac{dq(\hat{\omega_{l}})}{\omega_{l}}+\lambda=0 \label{14_1}
\end{eqnarray}
Now we subtract (\ref{14_1}) from (\ref{13_1}) and using the result from
Lemma 4.2 we have
\begin{eqnarray}
\sum_{j\neq k,j\neq l}&&\rho_{jk}^{*}
q(\hat{\omega_{j}})(\frac{dq(\hat{\omega_{l}})}{\omega_{l}}
-\frac{dq(\hat{\omega_{k}})}{\hat{\omega_{k}}}) \nonumber \\
&+&
\rho_{lk}^{*}\big(q(\hat{\omega_{k}})\frac{dq(\hat{\omega_{l}})}{\omega_{l}}\nonumber \\
&&~~~~~ - q(\hat{\omega_{l}})\frac{dq(\hat{\omega_{k}})}{\omega_{k}}\big)=0 \label{15}
\end{eqnarray}
The equation (\ref{15}) is trivially satisfied, if $\hat{\omega_{k}}=\hat{\omega_{l}}$. Another
possible root is $\hat{\omega}_{k}\neq\hat{\omega}_{l}>0$, where we have the following condition,
\begin{equation}
\sum_{j\neq k,j\neq l}\rho_{jk}^{*}q(\hat{\omega}_{j})=-\frac{\rho_{lk}^{*}[q(\hat{\omega}_{k})\frac{dq(\hat{\omega}_{l})}{\omega_{l}}-
q(\hat{\omega}_{l})\frac{dq(\hat{\omega}_{k})}{\omega_{k}}]}{\frac{dq(\hat{\omega}_{l})}{\omega_{l}}-\frac{dq(\hat{\omega}_{k})}{\hat{\omega_{k}}}}.\label{16}
\end{equation}
Substitute (\ref{16}) into (\ref{11_1}) we have the following
equation,
\begin{eqnarray}
L(\omega_{k},\omega_{l})&=&\|\y-\x\bfb\|_{2}^{2}\nonumber \\
&+&[2q_{k}+2q_{l}]\frac{\rho_{lk}^{*}[q(\hat{\omega}_{k})\frac{dq(\hat{\omega}_{l})}{\omega_{l}}-
q(\hat{\omega}_{l})\frac{dq(\hat{\omega}_{k})}{\omega_{k}}]}{\frac{dq(\hat{\omega}_{l})}{\omega_{l}}
-\frac{dq(\hat{\omega}_{k})}{\omega_{k}}}
\nonumber \\
&-&2\rho_{kl}^{*}q_{k}q_{l}+\lambda(|\omega_{k}|+|\omega_{l}|) \label{17}
\end{eqnarray}
Since $\hat{\omega}_{k}\neq\hat{\omega}_{l}$ is the optimal solution, then
$L(\hat{\omega}_{k},\hat{\omega}_{l})$ should be the minimum. Consider
another solution $S2=(\hat{\omega_{k}}+\hat{\omega_{l}},0)$. Since
$x_{k}\approx x_{l}$, then
$x_{k}\hat{\omega}_{k}+x_{l}\hat{\omega}_{l}=x_{k}(\hat{\omega}_{k}+\hat{\omega}_{l})+x_{l}\times
0$.  Also
$\lambda(\hat{\omega}_{k}+\hat{\omega}_{l})=\lambda(\hat{\omega}_{k}+\hat{\omega}_{l}+0)$.
The only difference between the solution
$\hat{\omega_{k}}\neq\hat{\omega_{l}}$ and the new solution $S2$ would be the
following,
\begin{eqnarray}
&&\rho_{kl}^*\bigg[\frac{q(\hat{\omega}_{k})^{2}\frac{dq(\hat{\omega}_{l})}{d\omega}
-q(\hat{\omega}_{l})^{2}\frac{dq(\hat{\omega}_{k})}{d\omega}}{\frac{dq(\hat{\omega}_{l})}{d\omega}-\frac{dq(\hat{\omega}_{k})}{d\omega}
}\nonumber \\
&&~~~~-\frac{q(\hat{\omega}_{k}+\hat{\omega}_{l})[q(\hat{\omega}_{k})\frac{dq(\hat{\omega}_{l})}{d\omega}
-q(\hat{\omega}_{l})\frac{dq(\hat{\omega}_{k})}{d\omega}]}{\frac{dq(\hat{\omega}_{l})}{d\omega}-\frac{dq(\hat{\omega}_{k})}{d\omega}
}\bigg]\label{18_1}
\end{eqnarray}
The equation (\ref{18_1}) takes a positive value by assumption
(\ref{A4})which implies that L$(\hat{\omega}_{k}+\hat{\omega}_{l},0)\textless
L(\hat{\omega}_{k},\hat{\omega}_{l})$, hence a contradiction. So
$x_{k}\approx x_{l} \Rightarrow \hat{\omega}_{k}\approx\hat{\omega}_{l}.$
Q.E.D.

\vspace{2cm}

\noindent \textbf{Proof of Theorem \ref{thm_LGC_group_effect}}: Suppose $\hat{\bfb}$ is the optimal solution with
$\hat{\omega}_{k},\hat{\omega}_{l}>0$. At the optimal point,
\scalebox{1.3}{$\frac{\partial L}{\partial \omega_{k}}=0$} and
\scalebox{1.3}{$\frac{\partial L}{\partial \omega_{l}}=0$}, so we have
\begin{eqnarray}
&&-2x_{k}^{T}(\y-\x\hat{\bfb})-2\frac{dq(\hat{\omega}_{k})}{\hat{\omega}_{k}}\sum_{j\neq k,j\neq l}
\rho_{jk}^{*}q(\hat{\omega}_{j}) \nonumber\\
&&~~~-2\rho_{lk}^{*}q(\hat{\omega}_{l})\frac{dq(\hat{\omega}_{k})}{\omega_{k}}+\lambda=0,
   ~~~and  \label{13_2}\\
&&-2x_{l}^{T}(\y-\x\hat{\bfb})-2\frac{dq(\hat{\omega}_{l})}{\omega_{l}}\sum_{j\neq k,j\neq l}\rho_{jl}^{*}q(\hat{\omega}_{j}) \nonumber\\
&&~~~-2\rho_{lk}^{*}q(\hat{\omega}_{k})\frac{dq(\hat{\omega}_{l})}{\omega_{l}}+\lambda=0 \label{14_2}
\end{eqnarray}
Subtract (\ref{14_2}) from (\ref{13_2}) and after some operations we get
the following equation,
\begin{equation}
\begin{split}
&\rho_{kl}^{*}(q_{l}\frac{dq_{k}}{\omega_{k}}-q_{k}\frac{dq_{l}}{\omega_{l}})=(x_{l}-x_{k})^{T}(y-X\omega)+
\\
&\sum_{j\neq k,j\neq l}q_{j}(\rho_{jl}^{*}\frac{dq_{l}}{\omega_{l}}-\rho_{jk}^{*}\frac{dq_{k}}{\omega_{k}}),
\end{split} \label{14_3}
\end{equation}
where,
\begin{equation*}
\begin{split}
&\mid \rho_{kl}^{*}(q_{l}\frac{dq_{k}}{\omega_{k}}-q_{k}\frac{dq_{l}}{\omega_{l}})|\leq |(x_{l}-x_{k})^{T}(y-X\omega)|
\\
&+ |\sum_{j\neq k,j\neq l}q_{j}(\rho_{jl}^{*}\frac{dq_{l}}{\omega_{l}}-\rho_{jk}^{*}\frac{dq_{k}}{\omega_{k}})|
\end{split}
\end{equation*}
 by Cauchy-Schwarz inequality,
\begin{eqnarray*}
|\rho_{kl}^{*}|
\times|(q_{l}\frac{dq_{k}}{\omega_{k}}-q_{k}\frac{dq_{l}}{\omega_{l}})|\leq
|(x_{l}-x_{k})^{T}|\times|(y-X\omega)|\\
+|[q_{j}]_{j\neq k,l}|\times|(\frac{dq_{l}}{\omega_{l}}-\frac{dq_{k}}{\omega_{k}})| \times |[\rho_{jl}^{*}-\rho_{jk}^{*}]_{j\neq k,l}|.
\end{eqnarray*}
Note that $[q_{j}]_{j\neq k,l},[\rho_{jl}^{*}-\rho_{jk}^{*}]_{j\neq
  k,l}$ are $1 \times (p-2)$ vectors. We have
$|(x_{l}-x_{k})^{T}|=\sqrt{x_{l}^{T}x_{l}+x_{k}^{T}x_{k}-2x_{l}^{T}x_{k}}=\sqrt{2(1-\rho_{lk})}$.
By assumption (\ref{A1}) q is a concave function, so we can say that
$|(\frac{dq_{l}}{\omega_{l}}-\frac{dq_{k}}{\omega_{k}})|$ is bounded by
$\lambda\frac{\pi}{2}$ (The derivative of $q$ at $0$). Substitute these
developments in above equation to get the upper bound. Similarly due to
concavity of q it is evident that
$|(q_{l}-q_{k})|\frac{dq_{k}}{\omega_{k}}\leq
|(q_{l}\frac{dq_{k}}{\omega_{k}}-q_{k}\frac{dq_{l}}{\omega_{l}})|$, if we
have $\omega_{k}\geq\omega_{l}$. Finally, we can say $|y-X\omega|\leq |y|$.
Q.E.D.

\section*{Related work}
In this section we review the existing feature selection algorithms $\&$
compare our proposed method with them. Lasso proposed by \cite{Tibshirani.1996} is widely used for feature
selection. Copula lasso prior is a multivariate extension of lasso prior
which accounts for the correlation between the features. In fact the `lasso
with Gauss copula prior' reduces to the lasso prior when the correlation
between the features is 0.

EN proposed by \cite{Yan.2011} also incorporates the correlation
between the features through eigen vectors of data covariance matrix. It is a
Bayesian hybrid model which discovers the correlated features through
the eigen information extracted from the data. EN \cite{Zou.Hastie.2005} uses a weighted combination of $l_{1}$ and
$l_{2}$ norms to encourage a grouping effect, where strongly correlated
variables tend to be in or out of the model together. However EN
does not use the correlation information embedded in the data in contrast
with copula prior. Copula function can be used to develop a multivariate
version of the EN to capture the correlation information between the
features. However in this paper we restrict ourselves to the LC prior.

Ordered weight $l_1$ (OWL) algorithm \cite{Zeng.2016} is also capable of
selecting correlated features. But it forces the features in the same group
to have the same coefficient value which introduces bias in the model.
\cite{Frank.1993} discuss a general $l_{q}$ penalty function on the model
parameters for $q > 0$. It is known as the Bridge estimator. Lasso
is a special case of Bridge estimator corresponding to $q=1$. For $q <
1$ the bridge penalty function becomes non convex.

\cite{Zellner.1986} introduced the $g$-prior. This prior replicates the
covariance structure of the data. However it cannot produce sparse solutions.
Multivariate laplace distribution in which covariance structure is identical
to data covariance serves many useful purposes. First it can identify the
correlated features due to its built-in correlation structure and secondly it
has the ability to produce sparse solutions. However handling multivariate
laplace distribution is computationally difficult, so we have used copula
techniques to develop the multivariate distribution function for lasso.

\section*{Acknowledgment}

Sourish Das's work was partially supported by the Infosys Foundation
grant to CMI.

\bibliographystyle{acm}
\bibliography{bibliography}

\end{document}